\documentclass [12pt]{article}
\usepackage{amsmath}
\usepackage{amsfonts}
\usepackage{epsfig}
\begin{document}
\def\b{\bar}
\def\d{\partial}
\def\D{\Delta}
\def\cD{{\cal D}}
\def\cK{{\cal K}}
\def\f{\varphi}
\def\g{\gamma}
\def\G{\Gamma}
\def\l{\lambda}
\def\L{\Lambda}
\def\M{{\Cal M}}
\def\m{\mu}
\def\n{\nu}
\def\p{\psi}
\def\q{\b q}
\def\r{\rho}
\def\t{\tau}
\def\x{\phi}
\def\X{\~\xi}
\def\~{\widetilde}
\def\h{\eta}
\def\bZ{\bar Z}
\def\cY{\bar Y}
\def\bY3{\bar Y_{,3}}
\def\Y3{Y_{,3}}
\def\z{\zeta}
\def\Z{{\b\zeta}}
\def\Y{{\bar Y}}
\def\cZ{{\bar Z}}
\def\`{\dot}
\def\be{\begin{equation}}
\def\ee{\end{equation}}
\def\bea{\begin{eqnarray}}
\def\eea{\end{eqnarray}}
\def\half{\frac{1}{2}}
\def\fn{\footnote}
\def\bh{black hole \ }
\def\cL{{\cal L}}
\def\cH{{\cal H}}
\def\cF{{\cal F}}
\def\cP{{\cal P}}
\def\cM{{\cal M}}
\def\ik{ik}
\def\mn{{\mu\nu}}
\def\a{\alpha}

\title{NEW ASPECTS OF THE PROBLEM OF THE SOURCE OF THE KERR SPINNING PARTICLE}

\author{Alexander Burinskii \\
Theor.Physics Laboratory, NSI, Russian Academy of Sciences,\\ B.
Tulskaya 52 Moscow 115191 Russia, email: bur@ibrae.ac.ru}

\date{Essay written for the Gravity Research Foundation 2010
Awards for Essays on Gravitation. (March 31, 2010) } \maketitle

\begin{abstract}
We consider development of the models of the source of the
Kerr-Newman (KN) solution and  new aspects related with the
obtained recently field model based on a domain wall bubble with
superconducting interior. The internal Higgs field regularizes the
KN solution,  expelling electromagnetic field from interior to the
boundary of bubble. The KN source forms a gravitating soliton,
interior of which is similar to oscillating solitons (Q-balls,
oscillons), while exterior is consistent with the KN solution. We
obtain that a closed Wilson loop appears on the edge of the
bubble,  resulting in quantization of angular momentum of the
regularized solutions. A new holographic interpretation of the
mysterious twosheetedness of the Kerr geometry is given.
\end{abstract}

\newpage
There are many evidences that black holes (BH) are  akin to
elementary particles \cite{Sen}.
 Carter obtained that the KN BH solution has $g=2$ as that of
 the Dirac electron, \cite{DKS}, and there followed a series of the
works on the problem of the source of  KN spinning particle, and
on the models of the KN electron consistent with gravity
\cite{Isr,Bur0,Ham,Lop,BurTwi,BurKN,BurAxi,BurSen,Dym,BurSuper}.
Since spin of electron is very high, the  horizons of KN solution
disappear, opening the naked singular ring which should be
replaced by a regular matter source. The regularized BH solutions
may be considered as gravitating solitons \cite{Shnir,Volk},  the
nonperturbative field solutions of  the electroweak sector of
standard model \cite{Volk,Dash} which may realize important bridge
between quantum theory and gravity.

Regularization of the KN solution represents a very old and hard
problem related with specific {\it twosheeted structure} of the
Kerr geometry \cite{BurAxi,BurSen,BurA}. The Kerr-Schild (KS) form
of metric
 is \be g_\mn=\eta _\mn + 2 H k_\m k_\n , \quad H=\frac {mr -
e^2/2}{r^2 + a^2 \cos ^2 \theta}, \label{ksm} \ee and
electromagnetic vector potential is \be A^\m_{KN} = Re \frac e
{r+ia \cos \theta} k^\m\label{ksGA} , \ee where $k^\m(x^\m)$ is
the null vector field tangent to the Kerr principal null
congruence (PNC) \cite{DKS}, $ k_\m dx^\m = dr - dt - a \sin ^2
\theta d\phi_K , $ and $\eta^\mn $ is the auxiliary Minkowski
metric with Cartesian coordinates $x^\m=(t,x,y,z),$ related with
the Kerr oblate spheroidal coordinates $r,\theta, \phi_K $  as
follows

\begin{figure}[ht]
\centerline{\epsfig{figure=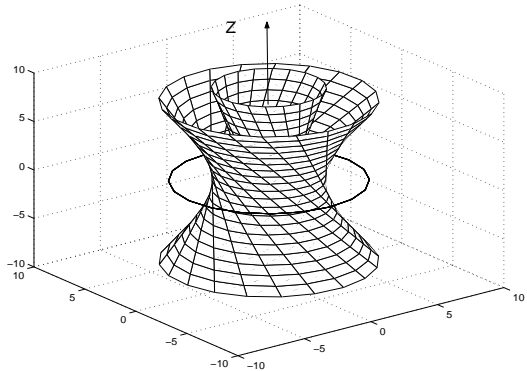,height=5cm,width=7cm}}
\caption{The Kerr singular ring and projection of the Kerr
congruence on auxiliary Minkowski background. }
\end{figure}

\be x+iy = (r + ia) \exp \{i\phi_K \} \sin \theta , \ z=r\cos
\theta \label{oblate},\ee where  $a=J/m$ is radius of the Kerr
singular ring, a branch line of the KN spacetime.

The coordinate $r$ covers the Kerr space-time twice, for $r>0$ and
for $r<0 ,$ forming the `positive' and `negative' sheets connected
analytically via disk $r=0, \cos\theta \le 1 ,$ see Fig.1. The
Kerr congruence $k_\m (x)$ covers the spacetime twice too: in the
form of ingoing rays $k^{\m (-)} $ falling on the disk $r=0,$ and
as outgoing rays $k^{\m (+)},$ for $r>0 ,$ which leads to
different metrics on the in- and out- sheets of the KN solution.

There are different models of the  KN source, and the we sketch
here some typical ones.
\begin{itemize}

\item[(1)]   Israel  \cite{Isr} (1968)  truncated  negative KN
sheet, replacing it by the {\it rotating disk}, $r=0 ,$ spanned by
the Kerr singular ring of the Compton radius $a=\hbar/2m.$

\item[(2)] In the suggested in \cite{Bur0} (1974) model of  {\it
"microgeon with spin"}, the Kerr singular ring was considered as a
waveguide for electromagnetic traveling waves generating the spin
and mass of the KN solution. Singular ring was interpreted as a
{\it closed `Alice' string} opening a gate to negative sheet of
spacetime \cite{BurKN,BurAxi,BurSen} (1995,2004,2008).

\item[(3)] L\'opez, \cite{Lop} (1984), generalized the Israel
model by introducing the ellipsoidal {\it bubble source} covering
the Kerr ring. The external KN solution matches with flat interior
along ellipsoidal boundary $r=r_0=e^2/2m ,$ forming an oblate
rigidly rotating bubble of the Compton size, with the thickness
$2r_0 = e^2/m ,$ equal to the classical size of electron.

\item[(4)] The {\it field and baglike models}: \cite{GG} (1974),
\cite{BurCas} (1989), \cite{BurBag} (2000), \cite{BEHM} (2002),
\cite{Dym} (2005), \cite{BurSuper} (2010) which are regular and
similar to gravitating solitons.

\end{itemize}
 The long-term development of the models of KN source resulted in
 the obtained recently field generalization of the L\'opez model,
 \cite{BurSuper}, based on a {\it domain wall bubble} interpolating
 between the external KN solution and internal superconducting
 pseudo-vacuum state.

{\it Gravitational sector of the model} is described by the metric
(\ref{ksm}) with the suggested in \cite{GG} function $H=f(r)/(r^2
+ a^2 \cos ^2\theta) ,$ which can describe the rotating metrics of
different types and match them smoothly,
\cite{GG,BurBag,BEHM}.\fn{The Kerr geometry is foliated into
rigidly rotating ellipsoidal layers $r=const.$ with angular
velocities $\Omega(r)= \frac a {a^2+r^2}.$} In particular, the
external KN metric, $f(r)=mr -e^2/2 ,$ matches with flat interior,
$f=0 ,$ along the ellipsoidal surface $ r =r_0 =e^2/2m .$

{\it Electromagnetic -- Higgs sector} is described by Higgs
Lagrangian, \cite{NO},

\be {\cal L}= -\frac 14 F_\mn F^\mn + \frac 12 \cD_\m \Phi \bar
\cD^\m \bar \Phi + V \label{L3} ,\ee where $\cD_\m =\nabla_\m +ie
A_\m ; $  $ F_\mn = A_{\m,\n} - A_{\n,\m} \ ;$  $\Phi =\Phi_0 \exp
\{i\chi \} , \ $ leading to

 \be \Box A_\m =I_\m = e |\Phi|^2 (\chi,_\m + e A_\m)
\label{Main}.\ee
\noindent Potential $V$ provides the phase
transition from external KN solution, where $\Phi =0 ,$ to
superconducting internal state with $|\Phi| = \Phi_0 > 0 .$

The bizarre Kerr coordinate $\phi_K ,$ (\ref{oblate}), is
inconsistent with the Higgs angular coordinate  $\phi.$ After
coordinate transformations  $\phi_K \to \phi$ , the potential
(\ref{ksGA}) takes the form
  \be A_\m dx^\m|_{r } =
\frac{-e r} {r^2 +a^2 \cos^2 \theta}[dt + a \sin ^2 \theta d\phi ]
+ \frac {2e r dr} {(r^2 +a^2)} .\label{APhi}\ee

\noindent It increases, approaching the boundary of bubble $r=r_0
= e^2/2m ,$ and in the equatorial plane, $ \cos\theta=0 ,$ it
reaches the magnitude \be A^{(edge)}_\m dx^\m = - \frac{2m} {e}[dt
+ a d\phi ] + \frac {2e r_0 dr} {(r_0^2 +a^2)}. \label{Aedge}\ee

The directions  $A^{(edge)}_\m$ in the equatorial plane are
tangent to the Kerr singular ring and form a closed loop at the
edge. The Wilson loop integral
  \be S^{(edge)}=\oint_{(edge)} A_\m(x)dx^\m =\oint
eA^{(edge)}_\phi d\phi=-4\pi ma =-4\pi J  \label{WL} \ee turns out
to be proportional to the KN angular momentum.

The  Higgs field $ \Phi(x) = \Phi_0 e^{i \chi(x)}$ expels the
electromagnetic field and current from the bulk of the
superconducting bubble, and we should set $I_\m=0 $ for $r<r_0 .$
It gives the internal solution \be \chi,_\m = - e A_\m^{(in)} , \
 \label{Ain}\ee as a full differential, and the second equation $\Box
A_\m^{(in)}=0$ is satisfied automatically. Taking the Higgs phase
in general form  $\chi =\omega t + n\phi + \chi_1(r),$ one obtains
from (\ref{Ain}) the internal solution

\be A_0^{(in)} =-\frac \omega e; \ A_\phi^{(in)} =-\frac n e; \
A_r^{(in)} = \chi_1'(r) /e  . \label{Ain}\ee
 Matching the edge field (\ref{Aedge}) with internal one, we obtain
 \be \omega=2m; \ J=ma=n/2; \ \chi_1(r) = - \ln (r^2 +a^2) \label{dchi}, \ee
 and therefore \be \Phi(x)= \Phi_0 \exp
\{i\chi \}= \Phi_0 \exp \{i 2m t - i \ln (r^2 +a^2) + i n \phi \}.
\label{Hin}\ee

\noindent Two important consequences follow from (\ref{dchi}):

i)The Higgs field forms a coherent vacuum state oscillating with
the frequency $\omega=2m ,$ similar to the soliton models of the
spinning Q-balls \cite{VolkWohn,Grah} and  bosonic stars
\cite{BosStar}.

ii) Angular momentum of the regular bubble source of the KN
solution is quantized, $J=ma=n/2, \ n=1,2,3,...$

The electromagnetic field and currents in a superconductor have a
`penetration depth'  $\delta \sim \frac 1{e|\Phi|} = 1/m_v$ ,
\cite{NO}. In our case it forms a thin surface layer, $r_0 -\delta
< r < r_0 ,$ in which the potential $A_\m$ differs from the
obtained solution (\ref{Ain}). Its deviation, $ A_\m^{(\delta)} =
A_\m - A_\m^{(in)} $ obeys the massive equation
 \be \Box A_\m^{(\delta)} =m_v^2  A_\m^{(\delta)} \ ,\label{skin}\ee
  which shows that a massive vector meson with mass $m_v=e |\Phi|$
  resides at the KN bubble and generates
 the circular current $I_\m=e m_v^2  A_\m^{(\delta)}$ concentrating at the
 edge of bubble close to  Wilson loop. There may also be a spectrum of
 such solutions, which supports the stringy version (2) of the KN source.

Although in the considered model the Kerr singular ring is removed
 and the internal space is flat, the used oblate coordinate
system still contains the harmless ringlike coordinate
singularity, and the KN twosheetedness has been survived. The
inner superconducting  state may be extended analytically to
negative sheet, $r<0 ,$  forming a flat  superconducting
pseudo-vacuum state, having the zero total energy density. We
arrive at a  {\it holographic interpretation} of the KN
twosheetedness \cite{BurPreQ} which turns out to be necessary for
quantum treatment. The negative sheet is considered as an
in-vacuum space, separated from the physical out-sheet by the
holographically dual boundary of the bubble. The necessity of such
separation was suggested in particular by Gibbons, \cite{Gib}, who
separated the curved spacetime $\cal M$  into two time-ordered
regions $\cal M_-$ and $\cal M_+$ associated with ingoing and
outgoing vacuum states $|0_->$ and $|0_+> .$   Similar prequantum
spacetime was introduced for black holes by `t Hooft et.al. in
\cite{SHW}: the two sheets of the KN space correspond to the `t
Hooft holographic correspondence, in which the source forms a
membrane, holographically dual to the bulk $\cal M=\cal M_-
\bigcup \cal M_+ ,$ \cite{BurPreQ}.  The mysterious problem of
twosheetedness of the KN space-time turns into its advantage
related with a holographic Kerr-Schild structure \cite{BurPreQ}
adapted for  quantum treatment \cite{Gib}.

The obtained solitonlike
 source of the KN solution represents a bubble filled by
coherently oscillating Higgs field, the  typical feature of the
other `oscillon' or `breather' soliton solutions. For parameters
of electron $a=\hbar/2m \sim 10^{22} ,$ and the bubble forms a
strongly oblated, rotating disk of the Compton radius, which
corresponds to the size of electron dressed by virtual photons. We
arrive at the conclusion that the obtained inner coherent
structure of the Compton region, as well as the adjoined Wilson
loop and circular current\fn{Note, that circular currents of the
Compton size were experimentally confirmed long ago in the low
energy absorbtion of the  $\gamma$-rays in aluminium
\cite{Comp}.}, should apparently be considered as integral parts
of the consistent with gravity solitonlike electron structure.

\end{document}